\documentclass[12pt,onecolumn,prc,showpacs,preprintnumber,amsmath,
floatfix,amssymb]{revtex4}
\usepackage{epsfig}
\usepackage{graphicx}
\usepackage{dcolumn}
\usepackage{bm}
\def\var{\mbox{\boldmath $\varepsilon$}}
\def\r{\mbox{{\bf  r}}}
\def\p{\mbox{\boldmath $p$}}
\def\q{\mbox{\boldmath $q$}}
\def\k{\mbox{\boldmath $k$}}
\def\t{\mbox{\boldmath $t$}}
\allowdisplaybreaks[4]

\begin{document}
\title{Inclusive electron scattering off $^{12}$C, ${}^{40}$Ca, and ${}^{40}$Ar:
  effects of the meson exchange currents }
\author{A.~V.~Butkevich and S.~V.~Luchuk}
\affiliation{Institute for Nuclear Research,
Russian Academy of Sciences, Moscow 117312, Russia\\}
\date{\today}
\begin{abstract}

The scattering of electrons on carbon, calcium, and argon targets are
analyzed using an approach that incorporates the contributions to the
electromagnetic response functions from the quasielastic (QE), inelastic
processes, and two-particle and two-hole meson exchange current
($\mathrm{2p}$-$\mathrm{2h}$ MEC).
This approach describes well the whole energy spectrum of data at very
different kinematics. It is shown that the accuracy of the $(e,e')$ cross
section calculations in the region between the QE and delta-resonance peaks,
where the $\mathrm{2p}$-$\mathrm{2h}$ MEC contribution reaches its maximum
value, depends on the momentum transfer $|\q|$ and at $|\q|>500$ MeV the
calculated and measured cross sections are in agreement within the experimental
uncertainties.

\end{abstract}

\maketitle

\section{Introduction}

The current~\cite{NOvA1, T2K} and future~\cite{DUNE, HK2T} long-baseline
neutrino experiments aim at measuring the lepton CP violation phase, improving
the accuracy of the value of the mixing angle $\theta_{23}$, and determing
neutrino mass ordering. To evaluate the oscillation parameters, the
probabilities of neutrino oscillations as functions of neutrino energy are
measured. The neutrino beams are not monoenergetic and have broad distributions
that range from tens of MeVs to a few GeVs. This is one of the problems in
achieving a high level of accuracy of the oscillation parameters measurements.

In this energy range, charged-current (CC) quasielastic (QE) scattering induced
by both one- and two-body currents and resonance production are the main
contributions to the neutrino-nucleus scattering. The incident neutrino energy
is reconstructed using calorimetric methods, which rely not only on the
visible energy measured in the detector, but also on the models of the
neutrino-nucleus interactions that are implemented in neutrino event
generators. In addition
to its role in the reconstruction of the neutrino energy, the neutrino-nucleus
scattering model is critical for obtain background estimates, and for correct
extrapolations of the near detector constraints to the far detector in
analyses aimed at determing the neutrino oscillation parameters.

The modeling of neutrino-nucleus interactions in the energy range 
$\varepsilon_{\nu} \approx 0.2$--5 GeV is one of the most complicated issues
facing neutrino oscillation experiments. The description of nuclear effects is
one of the largest sources of systematic uncertainties despite use of the near
detector for tuning the nuclear models employed in the neutrino events
generator.
A significant systematic uncertainty arises from the description of scattering
induced by the two body meson exchange currents (MEC), which may produce
two-particle and two-hole final states. Such excitations are induced by
two-body
currents, hence, they go beyond the impulse approximation scheme in which the
probe interacts with only a single nucleon and corresponds to the
$\mathrm{1p}$-$\mathrm{1h}$ excitations. A poor modelling of these MEC processes
leads to a bias in the reconstruction of neutrino energy and thereby to large
systematic uncertainties in the neutrino oscillation parameters~\cite{NOvA2}.

In recent years many studies have been presented to improve our knowledge on
lepton-nucleus scattering~\cite{Katori, Alvares, BAV1, BAV2, BAV3, Martini1,
  Martini2, Nieves1, Nieves2, BAV4, Martini3, Simo, Megias1,
  Megias2, Megias3, Rocco, BAV5, BAV6, Dolan, Gon1, Gon2, Gon3}. Approaches
which go beyond the impulse approximation were developed in
Refs.~\cite{Blund, Martini1, Martini2, Nieves1, Nieves2, Megias1, Megias2,
 Rocco, BAV5, Gon1}. As neutrino beams have broad energy distributions, 
various contributions to the cross sections can significantly overlap with
each other making it difficult to identify, diagnose and remedy shortcoming of
nuclear models. On the other hand, in electron-scattering the energy and
momentum transfer are known and therefore measurements in kinematic
ranges and on 
targets of interest to neutrino experiments give an opportunity to validate
and improve the description of nuclear effects. Electron beams can be used to
investigate physics corresponding to different interaction mechanisms, by
measuring the nuclear response at energy transfers varied independently from
three-momentum transfer. The neutrino detectors are typically composed of
scintillator, water, or argon. There is a large body of electron-scattering
data on carbon and calcium and only a few data sets available for scattering
on argon.

Weak interactions of neutrino probe the nucleus in a similar way as
electromagnetic electron interactions. The vector part of the electroweak
interaction can be inferred directly from electron-scattering and the
influence of nuclear medium is the same as in neutrino-nucleus scattering.
Precise electron-scattering data give unique opportunity to validate nuclear
model employed in neutrino physics. A model unable to reproduce electron
measurements cannot be expected to provide accurate predictions for neutrino
cross sections. So, the detailed comparison with electron
scattering data (semi-inclusive and inclusive cross sections and response
functions) is a necessary test for any theoretical models used to describe
of the lepton-nucleus interaction.

In this work we test a joint calculation of the QE,
$\mathrm{2p}$-$\mathrm{2h}$ MEC, and inelastic
scattering contributions (RDWIA+MEC+RES approach) on
carbon, calcium, and argon, using the relativistic distorted-wave impulse
approximation (RDWIA) \cite{Pick, Udias, JKelly} for quasi-elastic response
and meson exchange currents response functions for $\mathrm{2p}$-$\mathrm{2h}$
final states
presented in Ref.\cite{BAV4}. For calculation of inelastic contributions to the
cross sections we adopt
parameterizations for the single-nucleon inelastic structure functions given
in Refs.\cite{Bost1, Bost2}, which provide a good description of the resonant
structure in $(e,e')$ cross sections and cover a wide kinematic region.
We compare the RDWIA+MEC+RES predictions with the whole energy spectrum of
$(e,e')$ data, including the recent JLab data for electron scattering on carbon
and argon. We also perform a comparison and analysis of the calculated cross
sections and data at the momentum transfer that corresponds to the region
between the QE and $\Delta$-resonance peak, where the
$\mathrm{2p}$-$\mathrm{2h}$ response is
peaked. 

In Sec.II we brifly introduce the formalism needed for studying electron
scattering off nuclei with quasielastic, $\mathrm{2p}$-$\mathrm{2h}$ MEC, and
resonance production contributions. We also describe brifly the basic aspects
of the models used for the calculations. The results are presented and
discussed in Sec.III. Our conclusions are summarized in Sec.IV.

\section{Formalism of electron-nucleus scattering, RDWIA,
  $\mathrm{2p}$-$\mathrm{2h}$ MEC, and inelastic responses}

We consider the inclusive electron-nucleus scattering
\begin{equation}\label{qe:incl}
e(k_i) + A(p_A)  \rightarrow e'(k_f) + X                      
\end{equation}
in the one-photon exchange approximation. Here $k_i=(\varepsilon_i,\k_i)$ and
$k_f=(\varepsilon_f,\k_f)$ are the initial and 
final lepton momenta, $p_A=(\varepsilon_A,\p_A)$ is 
the initial target momentum, $q=(\omega,\q)$ is the momentum transfer carried by 
the virtual photon, and $Q^2=-q^2=\q^2-\omega^2$ is the photon virtuality. 

\subsection{Electron-nucleus cross sections}

In the inclusive reactions (\ref{qe:incl}) only the outgoing lepton is
detected and the differential cross section can be written as
\begin{equation}
\frac{d^3\sigma}{d\varepsilon_f d\Omega_f} =
\frac{\varepsilon_f}{\varepsilon_i}
 \frac{\alpha^2}{Q^4} L_{\mu \nu}{W}^{\mu \nu},
\end{equation}
where $\Omega_f=(\theta,\phi)$ is the solid angle for the electron momentum, 
$\alpha\approx 1/137$ is the fine-structure constant, $L_{\mu \nu}$ is the
lepton tensor, and $W^{\mu \nu }$ is the electromagnetic nuclear tensor.
In terms of the longitudinal $R_L$ and transverse $R_T$ nuclear response
functions the cross section reduces to
\begin{equation}
\frac{d^3\sigma}{d\varepsilon_f d\Omega_f} =
\sigma_M\big(V_LR_L + V_TR_T\big),
\end{equation}
where 
\begin{equation}
\sigma_M = \frac{\alpha^2\cos^2 \theta/2}{4\varepsilon^2_i\sin^4 \theta/2} 
\end{equation}
is the Mott cross section. The coupling coefficients 
\begin{subequations}
\begin{align}
V_L &= \frac{Q^4}{\q^4},
\\                                                                       
V_T &=\Bigl(\frac{Q^2}{2\q^2} + \tan^2\frac{\theta}{2}\Bigr),
\end{align}
\end{subequations}
are kinematic factors depending on the lepton's kinematics. The response
functions are given in terms of components of the hadronic tensors
\begin{subequations}
\begin{align}
R_L &=W^{00},
\\
R_T &=W^{xx}+W^{yy},                                  
\end{align}
\label{R}
\end{subequations}
and depend on the variables ($Q^2, \omega$) or ($|\q|,\omega$). They describe 
the electromagnetic properties of the hadronic system. The relations between
the response functions and cross sections for longitudinally $\sigma_L$ and
transversely $\sigma_T$ polarized virtual photons are
\begin{subequations}
\begin{align}
R_L &= \frac{K}{(2\pi)^2\alpha}\Bigl(\frac{\q^2}{Q^2}\Bigr)\sigma_L,
\\                                                                       
R_T &= \frac{K}{2\pi^2\alpha}\sigma_T,
\end{align}
\end{subequations}
where $K=\omega-Q^2/2m$ is the equivalent energy of a real photon needed to
produce the same final mass state and $m$ is the mass of nucleon. 

All the nuclear structure information and final state interaction effects 
(FSI) are contained in the electromagnetic nuclear tensor. It is given by
expression
\begin{eqnarray}
W_{\mu \nu } = \sum_f \langle X\vert                           
J_{\mu}\vert A\rangle \langle A\vert
J^{\dagger}_{\nu}\vert X\rangle,              
\label{W}
\end{eqnarray}
were $J_{\mu}$ is the nuclear electromagnetic current operator that connects
the initial nucleus state $|A\rangle$ and the final state 
$|X\rangle$. The sum is taken over the scattering states corresponding to all
of allowed asymptotic configurations. This equation is very 
general and includes all possible channels. Thus, the hadron tensor can be
expanded as the sum of the $\mathrm{1p}$-$\mathrm{1h}$ and
$\mathrm{2p}$-$\mathrm{2h}$, plus additional channels,
including the inelastic electron-nucleus scattering $W_{in}$: 
\begin{eqnarray}
  W^{\mu \nu } &=& W^{\mu \nu}_{1p1h} + W^{\mu \nu}_{2p2h} + W^{\mu \nu}_{in} \cdots  
\label{W_9}
\end{eqnarray}
The hadronic tensors $W_{1p1h}$, $W_{2p2h}$, and
$W_{in}$ determine, correspondingly, the QE, $\mathrm{2p}$-$\mathrm{2h}$ MEC,
and inelastic response functions. Therefore, the functions $R_i$
in~Eq.(\ref{R}) can be 
written as a sum of the QE ($R_{i,QE}$), MEC ($R_{i,MEC}$), and inelastic
response functions ($R_{i,in}$)
\begin{eqnarray}
R_i &=& R_{i,QE} + R_{i,MEC} +R_{i,in}         
\label{R_12}
\end{eqnarray}

\subsection{Model}

We describe genuine QE electron-nuclear scattering within the RDWIA
approach. This formalism is entirely based on the impulse approximation, namely
one body currents. In this approximation the nuclear current is written as a
sum of single-nucleon currents and nuclear matrix element in Eq.~(\ref{W})
takes the form 
\begin{eqnarray}\label{Eq12}
\langle p,B\vert J^{\mu}\vert A\rangle &=& \int d^3r~ \exp(i\t\cdot\r)
\overline{\Psi}^{(-)}(\p,\r)
\Gamma^{\mu}\Phi(\r),                                                     
\end{eqnarray}
where $\Gamma^{\mu}$ is the vertex function, $\t=\varepsilon_B\q/W$ is the
recoil-corrected momentum transfer, $W=\sqrt{(m_A + \omega)^2 - \q^2}$ is the
invariant mass and $\Phi$ and $\Psi^{(-)}$ are relativistic bound-state and
outgoing wave functions.

For electron scattering, we use the electromagnetic vertex
function for a free nucleon
\begin{equation}
\Gamma^{\mu} = F_V(Q^2)\gamma^{\mu} + {i}\sigma^{\mu \nu}\frac{q_{\nu}}
{2m}F_M(Q^2),                                                          
\end{equation}
where $\sigma^{\mu \nu}=i[\gamma^{\mu},\gamma^{\nu}]/2$, $F_V$ and
$F_M$ are the Dirac and Pauli nucleon form factors.
We use the approximation of Ref.~\cite{MMD} for the Dirac and Pauli nucleon
form factors and employ the de Forest prescription~\cite{deFor} and Coulomb
gauge for the off-shell vector current vertex $\Gamma^{\mu}$, because the
bound nucleons are off-shell.

In RDWIA calculations the independent particle shell model (IPSM) is assumed 
for the nuclear structure. In Eq.(\ref{Eq12}) the relativistic bound-state wave 
function for nucleons $\Phi$ are obtained as the 
self-consistent solutions of a Dirac equation, derived 
within a relativistic mean-field approach, from a Lagrangian containing 
$\sigma$, $\omega$, and $\rho$ mesons~\cite{Serot}. The nucleon 
bound-state functions were calculated by the TIMORA code~\cite{Horow} with the
normalization factors $S$ relative to full occupancy of the IPSM 
orbitals. For carbon an average factor $\langle S \rangle \approx 89\%$
is used, and for ${}^{40}$Ca and ${}^{40}$Ar the occupancy is
$\langle S \rangle \approx 87\%$ on average. These estimations of the
depletion of the hole state follows from the RDWIA analysis of
${}^{12}$C$(e,e'p)$~\cite{Dutta, Kelly1} and
${}^{40}$Ca$(e,e'p)$~\cite{BAV4}.  In this work we assume that the source of
the reduction of the $(e,e'p)$ spectroscopic factors with respect to the mean
field values are the $NN$ short-range and tensor correlations in the ground
state, leading to the appearance of the high-momentum and high-energy
component in the nucleon distribution in the target. 

In the RDWIA, final state interaction effects for the outgoing nucleon are 
taken into account. The distorted-wave function of the knocked out nucleon
$\Psi$ is evaluated as a solution of a Dirac equation containing a
phenomenological relativistic optical potential. This potential consists of a
real part, which describes the rescattering of the ejected nucleon and an
imaginary part for the absorption of it into unobserved channels. The EDAD1
parameterization~\cite{Cooper} of the relativistic optical potential for carbon
and calcium was used in this work.
A complex optical potential with a nonzero imaginary part generally produces
an absorption of the flux. However, for the inclusive cross section, the total
flux must be conserved. The inclusive responses (i.e., no flux lost) can
be handled by simply removing the imaginary terms in the potential. This yields
results that are almost identical to those calculate via relativistic Green's
function approach~\cite{Meucci1, Meucci2} and Green's function Monte Carlo
method~\cite{Rocco2} in which the FSI effects are treated by means of complex
potential and total flux is conserved.

The inclusive cross sections with the FSI effects, taking into account the
$NN$ correlations were calculated using the method proposed in Ref.~\cite{BAV1}
with the nucleon high-momentum and high-energy distribution from
Ref.~\cite{Atti} renormalized to value of 11\% for carbon and of 13\%
for calcium and argon. The contribution of the $NN$-correlated pairs is
evaluated in the impulse approximation, i.e., the virtual photon couples to only
one member of the $NN$ pair. It is a one-body current process that leads to the
emission of two nucleons ($\mathrm{2p}$-$\mathrm{2h}$ excitation). 

The evaluation of the $\mathrm{2p}$-$\mathrm{2h}$ MEC contributions is
performed within the
relativistic Fermi gas model~\cite{Pace, Simo}. The short-range
$NN$-correlations and FSI effects were not considered in this approach.
The elementary hadronic tensor $W^{\mu \nu}_{2p2h}$ is given by the bilinear
product of the matrix elements of the two-body electromagnetic MEC. Only
one-pion exchange is included. 
The two-body current operator is obtained from the electroweak pion 
production amplitudes for the nucleon~\cite{Her} with the coupling a second 
nucleon to the emitted pion. The two-body electromagnetic current is the sum of
seagull, pion-in-flight, and Delta-pole currents. The seagull terms are
associated with the interaction of the virtual proton at the $NN\pi$ vertex,
whereas the pion-in-flight operator is referred to the direct interaction of
photon with the virtual pion. 
The $\Delta$ peak is the main contribution to the pion 
production cross section. However, inside the nucleus $\Delta$ can also decay 
into one nucleon that rescatters producing two-nucleon emission without pions. 
As a result, the MEC peak is located in the dip region between the QE and 
Delta peaks, i.e., the invariant mass of the pion-nucleon pair
 $W^2=(q+p_A)^2=m^2 + 2m\omega -Q^2$  
varies in the range $(m_{\pi}+m)\leq W \leq 1.3-1.4$ GeV, where $m_{\pi}$ is the
mass of pion.

The exact evaluation of the $\mathrm{2p}$-$\mathrm{2h}$ hadronic tensor in a
fully relativistic way performed in Refs.~\cite{Pace, Simo} is highly
non-trivial.
In the present work we evaluate the electromagnetic MEC response functions 
$R_{i, MEC}$ of electron scattering on carbon using accurate parameterizations
of the exact MEC calculations. The $\mathrm{2p}$-$\mathrm{2h}$ MEC
contributions for ${}^{40}$Ca and
${}^{40}$Ar were calculated using the parameterization for ${}^{12}$C rescaled
for calcium and argon according to Ref.~\cite{Megias4}. The parameterization
form employed for the different electroweak responses is the function of
$(\omega,|\q|)$ and valid in the range of momentum transfer
$|\q|=200$--2000 MeV. The expressions for the fitting parameters are
described in detail in Refs.~\cite{Megias2, Megias3, MeAm}.

Finally, the inelastic response functions $R_{i,in}$ were calculated using the
parameterization for the neutron~\cite{Bost1} and proton~\cite{Bost2}
structure functions. This approach is based on an empirical fit to describe the
measurements of inelastic electron-proton and electron-deuteron cross sections
in the kinematic range of four-momentum transfer $0 < Q^2 <8$ GeV$^2$ and
final state invariant mass $1.1 < W_x < 3.1$ GeV, thus starting from pion
production region to the highly-inelastic region. These fits are constrained by
the high precision longitudinal $\sigma_L$ and transverse $\sigma_T$
separated cross section measurements and provide a good description of the
structures seen in inclusive $(e,e')$ cross sections.
 
\section{Results and analysis}

Before providing reliable predictions for neutrino scattering, any model must be
validated by confronting it with electron scattering data. The
agreement between the model's predictions and data in the vector sector of
electroweak interaction gives us confidence in the extension of this
phenomenological approach and its validity at least in the vector sector of
the electroweak interaction.

To test the RDWIA+MEC+RES approach we calculated the double-differential
inclusive ${}^{12}$C$(e,e')$, ${}^{40}$Ca$(e,e')$, and ${}^{40}$Ar$(e,e')$ 
cross sections as functions of the energy transfer to the nucleus.
\begin{figure*}
  \begin{center}
    \includegraphics[height=19cm,width=19cm]{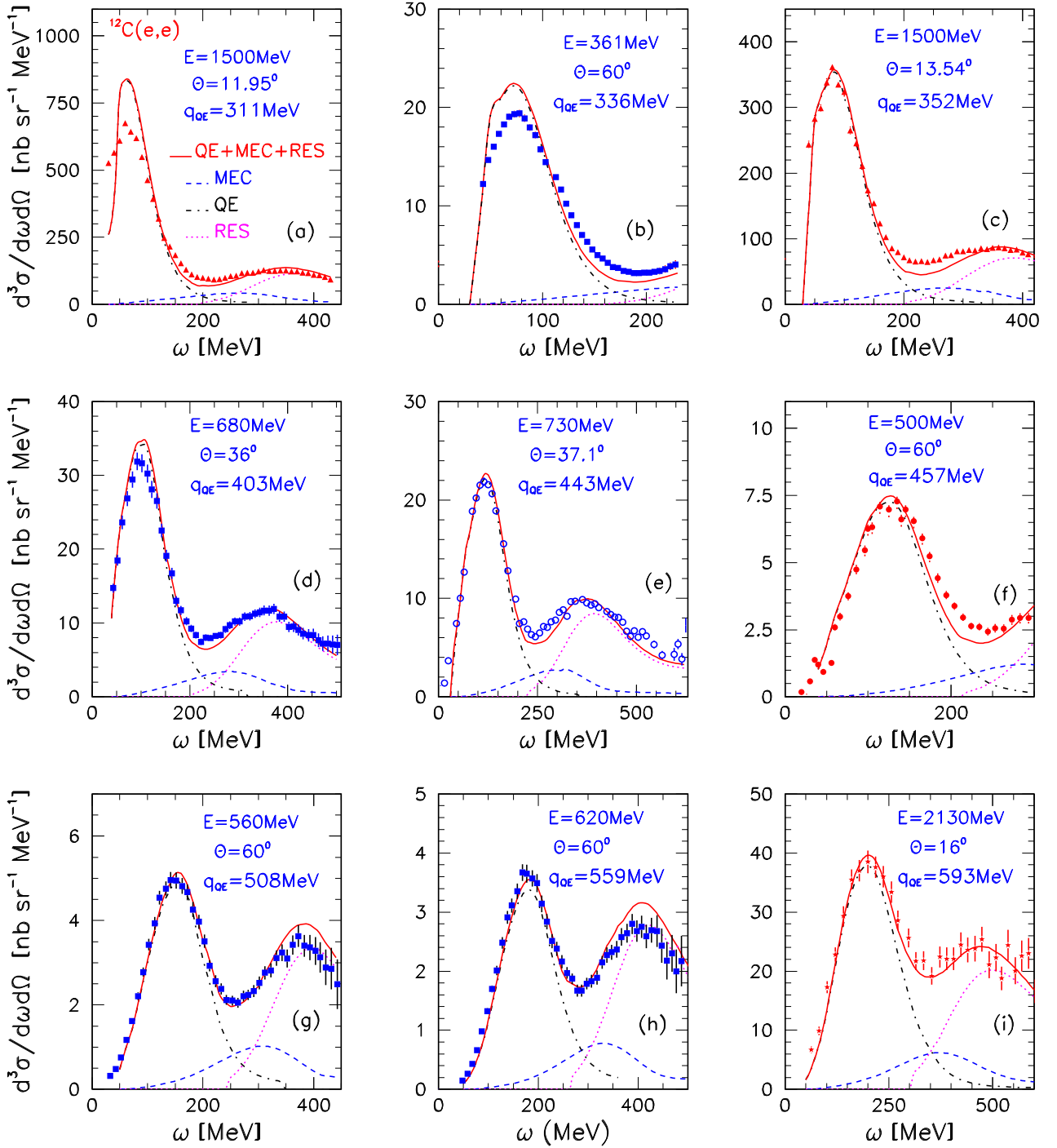}
  \end{center}
  \caption{\label{Fig1} The ${}^{12}$C$(e,e')$ double differential cross
    sections as functions of energy transfer $\omega$ compared with the
    RDWIA+MEC+RES
    predictions. The data are from Ref.~\cite{Baran} (filled triangles), Ref.~
    \cite{Barreau} (filled squares), Ref.~\cite{Connel} (open circles), Ref.
    ~\cite{Whitney}(filled circles), Ref.~\cite{Benhar1, Benhar2} (stars).
}
\end{figure*}
\begin{figure*}
  \begin{center}
    \includegraphics[height=19cm,width=19cm]{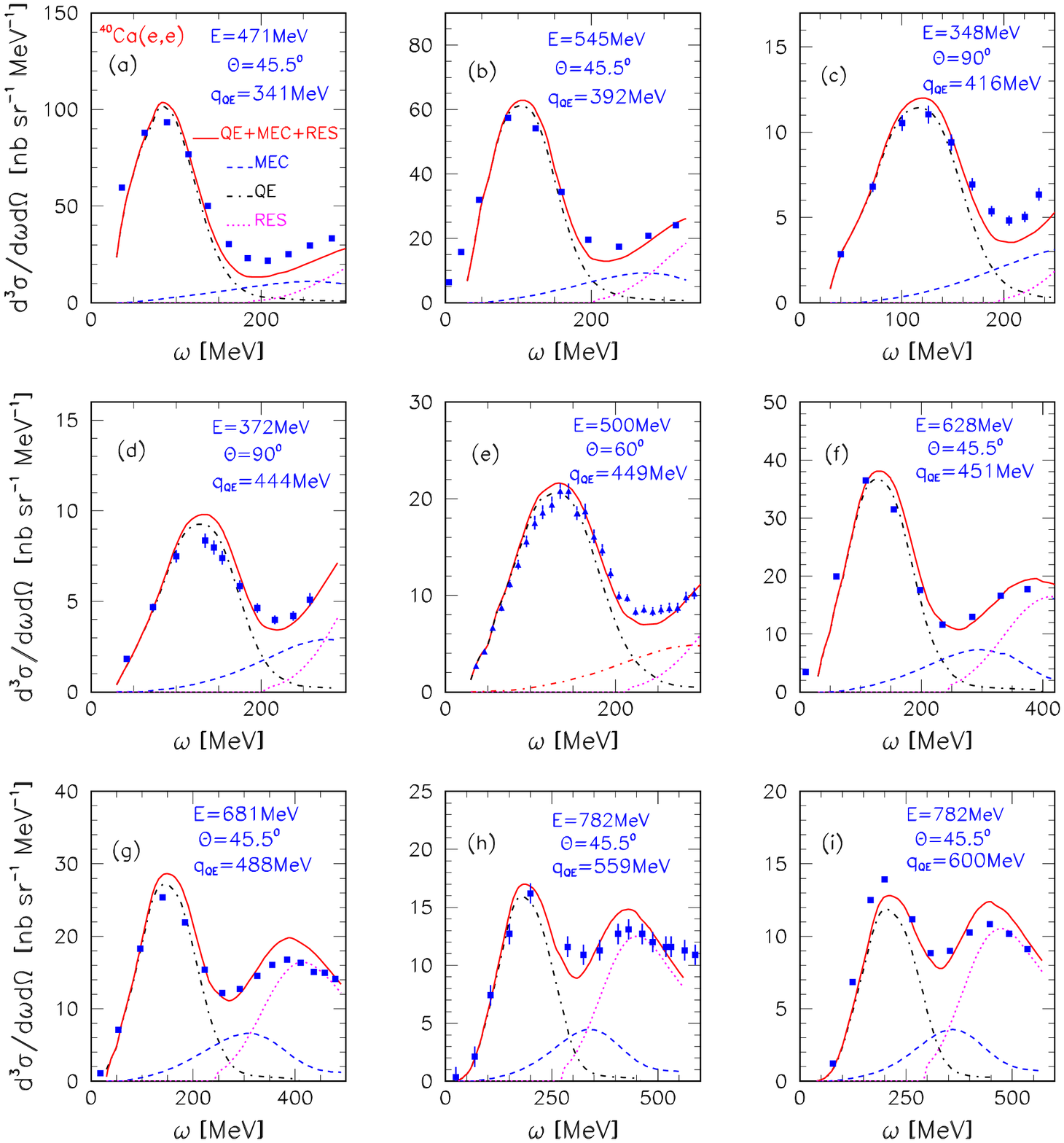}
  \end{center}
  \caption{\label{Fig2} The ${}^{12}$Ca$(e,e')$ double differential cross
    sections as functions of energy transfer $\omega$ compared with the
    RDWIA+MEC+RES predictions. The data are from Ref.~\cite{William} (filled
    squares) and Ref.~\cite{Whitney} (filled triangles).}
\end{figure*}
Results for
carbon and calcium are shown in Figs.~\ref{Fig1} and \ref{Fig2}, respectively,
and compared with data from Refs.~\cite{Barreau, Whitney, Connel, Sealock,
  Baran, William, Benhar1, Benhar2}. Each panel corresponds to the fixed values
 of the incident electron energy $E$ and scattering angle $\theta$.
 The kinematical coverage includes both quasielastic peak, dip region, and
 extends to the region of the delta-production peak.

 In Figs.~\ref{Fig1} and \ref{Fig2} we show the separate contributions to the
 inclusive cross section from QE (dot-dashed line),
 $\mathrm{2p}$-$\mathrm{2h}$  MEC (dashed line), and inelastic (dotted
 line) processes. The total contribution is presented by a solid line.
The panels have been ordered according to the corresponding value for the
momentum transfer at the quasielastic peak $q_{QE}$. This corresponds to the
value of $|\q|$ where the maximum in the QE peak appears. The $q_{QE}$ 
runs from $\approx$ 310 MeV to $\approx$ 590 MeV for carbon and
$340 < q_{QE} < 600$ MeV for calcium.

The systematic analysis presented in Figs.~\ref{Fig1} and \ref{Fig2} shows that
the RDWIA+MEC+RES approach leads to a good description of the whole set of
$(e,e')$ data, validating the reliability of our predictions. The positions,
widths, and heights of the QE peak are reproduced by the model within the
experimental errors, taking into account not only the QE domain but also the
contributions given by the $\mathrm{2p}$-$\mathrm{2h}$ MEC and inelastic terms.
Notice that the dip region is also successfully reproduced by the theory.
Only at the lower value of $q_{QE}<340$ MeV the theoretical predictions for
carbon overestimate data by 30-50\% at the QE peak, and this should be expected
since this is the region where the impulse approximation conditions may not be
satisfied and collective nuclear effects are important. 

The agreement between theory and data in the inelastic region also is good
within the experimental uncertainties. The inelastic part of the cross section
is dominanted by the $\Delta$ peak that contributes to the transverse response
function. In particular, $\omega_{QE}=\sqrt{|\q|^2+m^2}-m$ corresponds roughly
to the center of the quasielastic peak, and 
$\omega_{\Delta}=\sqrt{|\q|^2+m^2_{\Delta}}-m$ to the 
$\Delta$-resonance [$m_{\Delta}$ is the mass of $\Delta(1232)$]. When the
momentum transfer is not too high these regions are clearly separated in data
\begin{equation}
\Delta \omega=\omega_{\Delta}-\omega_{QE}= \frac{(m^2_{\Delta}-m^2)}
{\sqrt{|\q^2|+m^2} + \sqrt{|\q^2|+m^2_{\Delta}}},                   
\end{equation}
allowing for a test of theoretical models for each specific process. On the
other hand, for increasing values of the momentum transfer the peaks
corresponding to the $\Delta$ and QE domains become closer, and their overlap
increases significantly. In this case only the comparison with a complete model
including inelastic processes is meaningful.
\begin{figure*}
  \begin{center}
    \includegraphics[height=18cm,width=18cm]{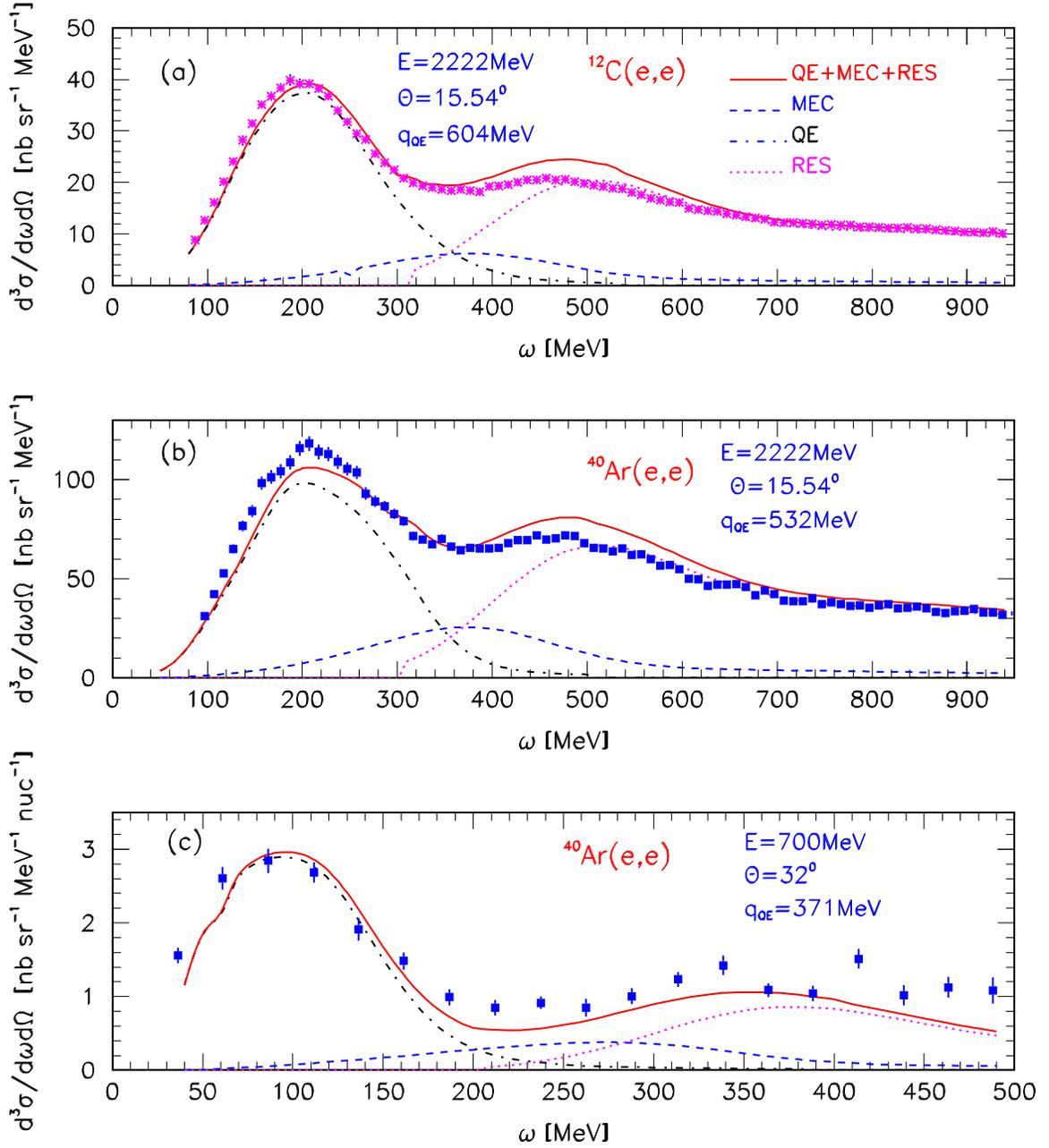}
  \end{center}
  \caption{\label{Fig3} The ${}^{12}$C$(e,e')$ (a) and ${}^{40}$Ar$(e,e')$ (b),
    (c) double differential cross sections of carbon and argon from
    Refs.~\cite{Dai1, Dai2} vs energy transfer $\omega$, compared with the
    RDWIA+MEC+RES prediction. The beam energy is $E=2.222$ GeV and scattering
    angle $\theta=15.541^{\circ}$. For completeness,
    data for electron scattering off argon at $E=700$ MeV and $\theta=32^{\circ}$
    from Ref.~\cite{William} are also shown (c). As shown in the key, the
    separate QE, $\mathrm{2p}$-$\mathrm{2h}$ MEC, and inelastic contributions
    are presented.}
\end{figure*}

In addition to the previous analysis, we have also tested the validity of the
RDWIA+MEC+RES approach through the analysis of the recent JLab data~\cite
{Dai1, Dai2} for inclusive electron scattering data on carbon and argon at
incident electron energy $E=2.222$ GeV and scattering angle
$\theta=15.54^{\circ}$. As observed in Fig.~\ref{Fig3}, the agreement between
theory and data is very good over most of the energy spectrum, with some minor
discrepancy seen only at the $\Delta$-resonance peak. For completeness, we also
present in this figure the electron-argon scattering spectrum measured at
the beam energy $E=700$ MeV and scattering angle  $\theta=32^{\circ}$~\cite
{Anghi}. Note that the $\mathrm{2p}$-$\mathrm{2h}$ MEC response, peaked in the
dip region between the QE and $\Delta$ peaks is essential to reproduce the data.

\begin{figure*}
  \begin{center}
    \includegraphics[height=14cm,width=14cm]{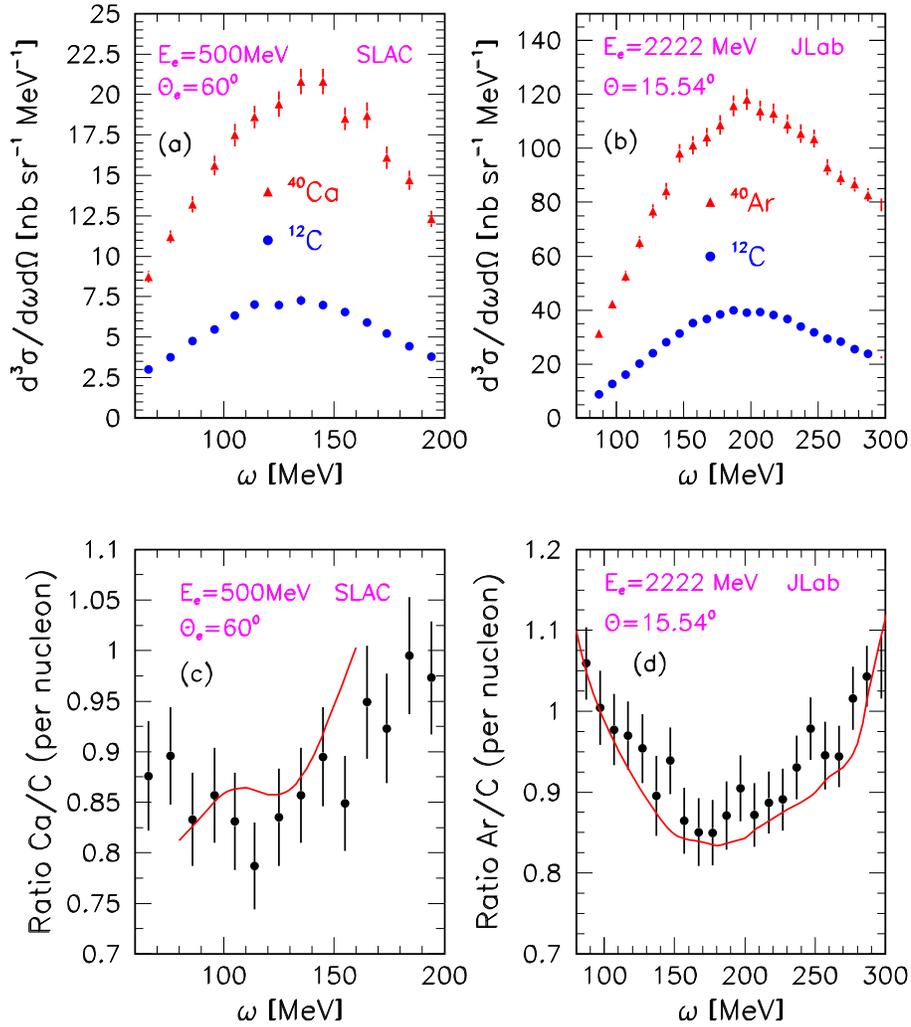}
  \end{center}
  \caption{\label{Fig4} The inclusive cross sections (a) and (b) and
    per nucleon cross section ratios $Ca/C$ (c) and $Ar/C$ (d) as
    functions of energy transfer $\omega$ for electron scattering on
    ${}^{12}$C, ${}^{40}$Ca, and ${}^{40}$Ar. Data for ${}^{40}$Ca and ${}^{12}$C
    (a) are from Ref.~\cite{Whitney} for electron beam energy $E=500$ MeV and
    scattering angle $\theta=60^{\circ}$. Data for ${}^{40}$Ar and ${}^{12}$C (b)
    are from Ref.~\cite{Dai1,Dai2} for $E=2222$ MeV and $\theta=15.54^{\circ}$.
    The solid line is the result of the RDWIA+MEC+RES calculation.} 
\end{figure*}

In the SLAC experiment~\cite{Whitney} the inclusive cross sections
$d\sigma/d\var d\Omega$ for electron scattering on ${}^{12}$C and ${}^{40}$Ca
were measured in the same kinematical conditions, i.e., at incident electron
energy $E=500$ MeV and $\theta=32^{\circ}$. Using the SLAC and JLab data we
estimated the measured
$(Ca/C)=(d\sigma^{Ca}/d\var d\Omega)_{nucl}/(d\sigma^{C}/d\var d\Omega)_{nucl}$
and 
$(Ar/C)=(d\sigma^{Ar}/d\var d\Omega)_{nucl}/(d\sigma^{C}/d\var d\Omega)_{nucl}$
ratios, where the differential cross sections
$(d\sigma^{i}/d\var d\Omega)_{nucl}$ are scaled with the number of nucleons in
the targets. Figure~\ref{Fig4} shows the measured ratios as functions of energy
transfer as compared to the RDWIA+MEC+RES calculations in the QE peak region.
The calculated $(Ca/C)$ ~\cite{BAV4} and $(Ar/C)$ ratios agree with data where
the observed effects of $\approx 15$\% in the QE peak region is higher than
experimental errors. In Ref.~\cite{BAV4} it was shown that the ground-state
properties of these nuclei and FSI effects give the dominant contributions to
the difference between the ${}^{12}$C and ${}^{40}$Ca(${}^{40}$Ar) differential
cross section per nucleon. The difference between the results for the carbon and
argon targets is relevant in the context of Monte Carlo simulation for the DUNE
neutrino oscillation experiment, where liquid argon and scintillator detectors
are planned to be used as near detectors. 

The agreement between theory and data in the dip region also gives us a
confidence in the reliability of our calculations of the MEC effects. In this
region the
contributions emerge from the QE, $\mathrm{2p}$-$\mathrm{2h}$ MEC, and inelastic domains and can
significantly overlap with each other making it difficult to experimentally
separate the different reaction channels, for instance, the QE and two-nucleon
knockout responses. Therefore, the comparison with data in this region can be
considered to be a critical test for the validity of the RDWIA+MEC+RES
approach, and particularly, the description of the $\mathrm{2p}$-$\mathrm{2h}$
MEC contribution that
reaches its maximum value here. We can consider the difference
between the calculated and measured cross sections observed at the maximum of
the $\mathrm{2p}$-$\mathrm{2h}$ MEC contribution as a conservative estimate of
the accuracy of the MEC response calculation in the vector sector of the
electroweak interaction.

The electron scattering cross sections on carbon and calcium with
scattering angle $\theta<60^{\circ}$, corresponding to the kinematic of the
neutrino oscillation experiments were analyzed. The ${}^{12}$C$(e,e')$ data were
divided into two sets with electron energies $0.4 \leq E  \leq 1.2$ GeV and
$1.5 \leq E \leq 3.5$ GeV, that approximately corresponds to neutrino energies
of the T2K (low energy) and NOvA (high energy) experiments.

We calculated
$R^{i}_{dip}=(d\sigma^i/d\var d\Omega)_{cal}/(d\sigma^i/d\var d\Omega)_{data}$
ratios at the momentum transfer $|\q|_{dip}$ that corresponds to the minimum
of the measured cross section, where $(d\sigma^{i}/d\var d\Omega)_{cal}$ and  
$(d\sigma^{i}/d\var d\Omega)_{data}$ are calculated and measured cross sections,
correspondingly, for electron scattering off carbon ($i$=C) and calcium
($i$=Ca).
\begin{figure*}
  \begin{center}
    \includegraphics[height=14cm,width=14cm]{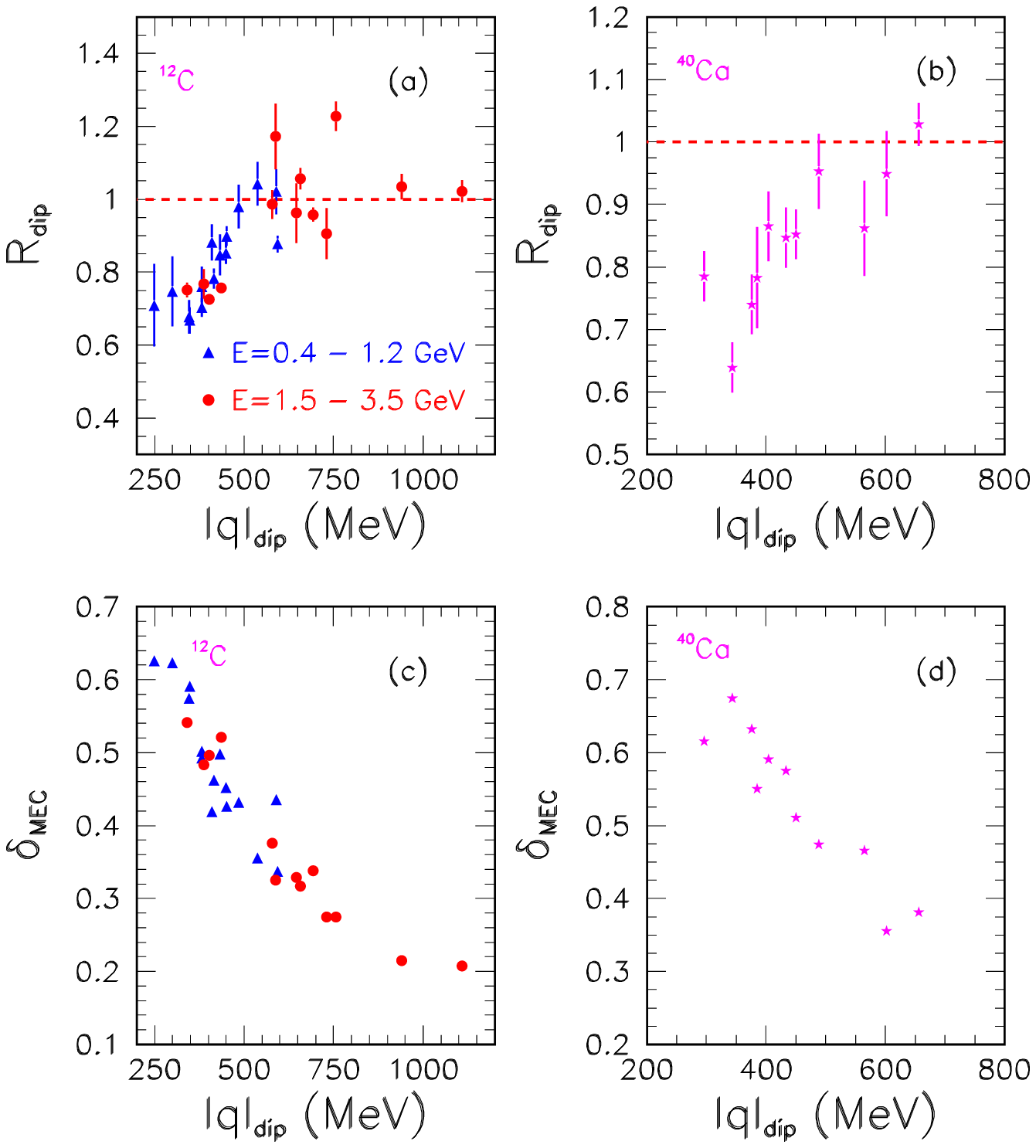}
  \end{center}
  \caption{\label{Fig5} Ratio $R^C_{dip}$ for carbon (a) and $R^{Ca}_{dip}$ for
  calcium (b) as a function of $|\q|_{dip}$. For carbon, data are from
  Refs.~\cite{Benhar1, Benhar2} and for calcium from Ref.~\cite{William}.
  The ratios $R^{C}_{dip}$ are shown for the two ranges of the incident electron
  energy $E=0.4 - 1.2$ GeV (filled triangles) and $E=1.5 - 3.5$ GeV (filled
  circles). The results of the RDWIA+MEC+RES calculation of the
  $\mathrm{2p}$-$\mathrm{2h}$ MEC contributions $\delta_{MEC}$ vs $|\q|_{dip}$
  for electron scattering on carbon (c) and calcium (d). As shown in the key
  the contributions for carbon are shown for $E=0.4 - 1.2$ GeV and
  $E=1.5 - 3.5$ GeV.} 
\end{figure*}
The values of $|\q|_{dip}$ running from $\approx 250$ MeV to
$\approx 1100$ MeV for carbon and $340 \leq |\q|_{dip} \leq 660$ MeV for
calcium. We also calculated the $\mathrm{2p}$-$\mathrm{2h}$ MEC contributions
to the $(e,e')$ differential cross sections, i.e.,
$\delta_{MEC}=(d\sigma/d\var d\Omega)_{MEC}/(d\sigma/d\var d\Omega)$
ratios, where the $(d\sigma/d\var d\Omega)_{MEC})$ is the
$\mathrm{2p}$-$\mathrm{2h}$ MEC
differential cross sections for electron scattering off nuclei.
\begin{figure*}
  \begin{center}
    \includegraphics[height=14cm,width=14cm]{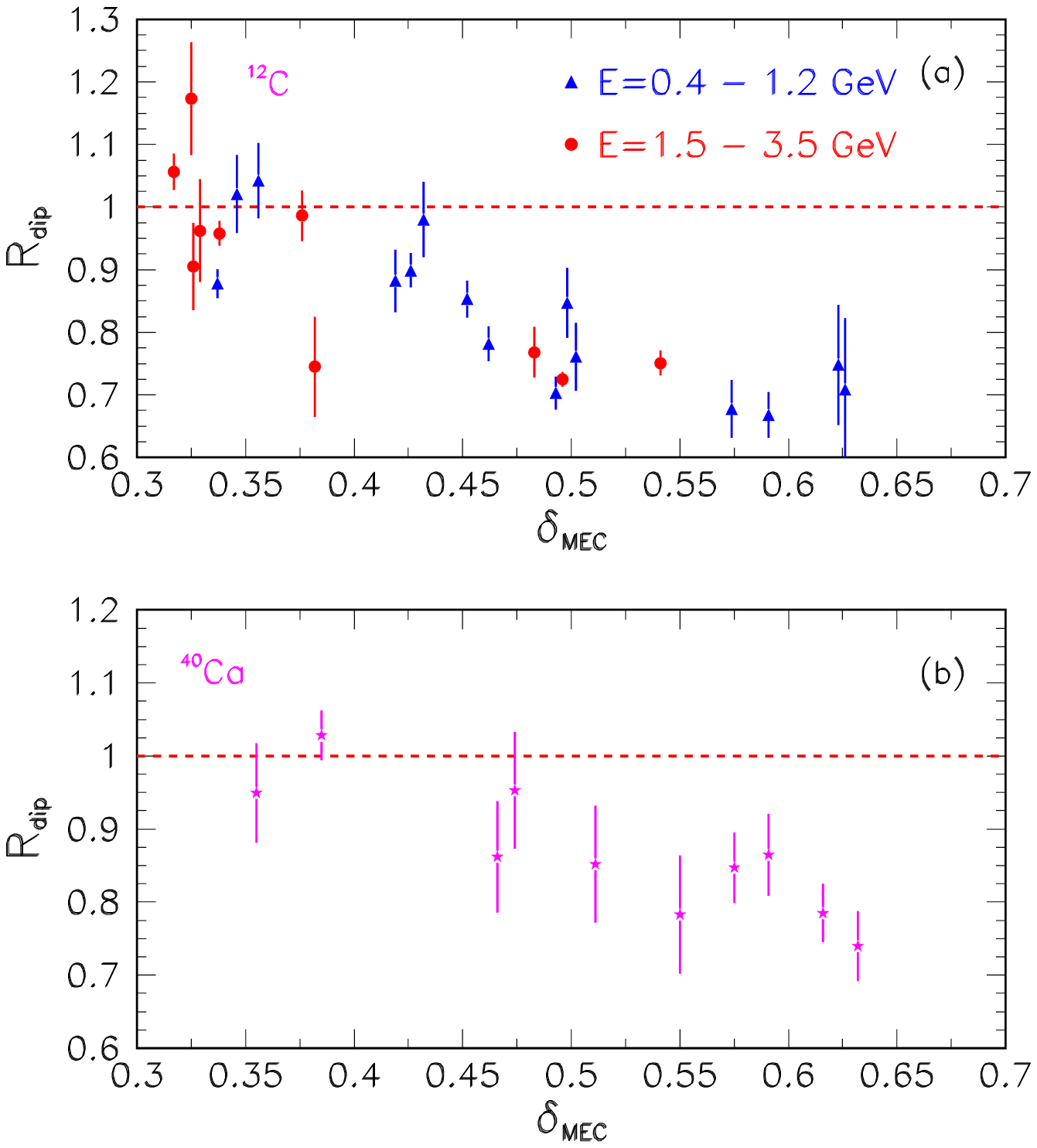}
  \end{center}
  \caption{\label{Fig6} Ratio $R^C_{dip}$ for carbon (a) and $R^{Ca}_{dip}$ for
    calcium (b) as a function of the $\mathrm{2p}$-$\mathrm{2h}$ MEC contribution $\delta_{MEC}$,
    calculated in the RDWIA+MEC+RES approach. The ratios for carbon are
    shown for the incident electron energies $E=0.4 - 1.2$ GeV (filled
    triangles) and $E=1.5 - 3.5$ GeV (filled circles).} 
 \end{figure*}
Figure~\ref{Fig5} shows the ratios $R^i_{dip}$ and $\delta_{MEC}$ as functions
of $|\q|_{dip}$. The result presented in Fig.~\ref{Fig5}(a) demonstrates that
the $R^C_{dip}$ ratio increases with $|\q|_{dip}$ from 0.7 at
$|\q|_{dip} \approx 250$ MeV to $\approx 1$ at $|\q|_{dip} \approx 500$ MeV and
does not depend on electron energy. At $|\q|_{dip} > 500$ MeV the calculated
and measured cross sections are in good agreement within the
experimental errors. On the other hand the contribution $\delta_{MEC}$
[Fig.~\ref{Fig5}(c)] reduces with momentum transfer from 0.65 at
$|\q|_{dip}\approx 250$ MeV to 0.42 at $|\q|_{dip}\approx 500$, and up to
0.2 at $|\q|_{dip}\approx 1000$ MeV and also does not depend on
the electron energy.
The ratio $R^{Ca}_{dip}$ [Fig.~\ref{Fig5}(b)] shows a similar dependence on
$|\q|_{dip}$, i.e., $R^{Ca}_{dip}$ increases with $|\q|_{dip}$ from 0.7 at
$|\q|_{dip}= 350$ MeV to $\approx 1$ at $|\q|_{dip} > 500$ MeV. The
$\mathrm{2p}$-$\mathrm{2h}$ MEC contribution [Fig.~\ref{Fig5}(d)] decreases
with momentum transfer from 0.68 at $|\q|_{dip}=300$ MeV, and up to 0.38 at
$|\q|_{dip}=600$ MeV.     

The $R^{i}_{dip}$ ratios for carbon (upper panel) and calcium (lower panel) are
shown in Fig.~\ref{Fig6} as functions of $\delta_{MEC}$. The figure shows that
the $R^i_{dip}\approx 1$ up to $\delta_{MEC} \approx 0.45$ and then is reduced
with $\delta_{MEC}$ to $\approx 0.8$ at $\delta_{MEC} \approx 0.6$. Thus, the
contribution of the $\mathrm{2p}$-$\mathrm{2h}$ MEC decreases with momentum
transfer and the
accuracy of the inclusive cross section calculated within the RDWIA+MEC+RES
approach in the dip region improves with $|\q|_{dip}$ from 35\% at
$|\q|_{dip}\approx 250$ MeV ($|\q|_{dip}\approx k_F$) to 10\% at
$|\q|_{dip}\geq 500$ MeV ($|\q|_{dip} \geq 2k_F$), where $k_F$ is the Fermi
momentum. We can use this estimation as conservative estimate of the accuracy
of the $\mathrm{2p}$-$\mathrm{2h}$ MEC response calculation in the vector sector of the electroweak
interaction.
 
\section{Conclusions}

In this article, we studied the quasielastic, $\mathrm{2p}$-$\mathrm{2h}$ MEC,
and inelastic
electron scattering on carbon, calcium, and argon targets in the RDWIA+MEC+RES
approach. This approach was extended to the whole energy spectrum, incorporating
the contributions coming from the QE, inelastic and $\mathrm{2p}$-$\mathrm{2h}$
 meson exchange
currents. In calculation of the QE cross sections within the RDWIA, the effects
of FSI and short-range $NN$-correlations in the target ground state were
taking into account. An accurate parameterization of the exact MEC calculations
of the nuclear response functions was used to evaluate the MEC response. The
inelastic response functions were calculated using the parameterization for
the neutron and proton structure functions. These functions were obtained from
the fit of the measured inelastic electron-proton and electron-deuteron cross
sections.

The present approach is capable of reproducing successfully the whole energy
spectrum of $(e,e')$ data at very different kinematics, including the recent
JLab data for inclusive electron scattering on carbon and argon. It was shown
that the measured and calculated in the RDWIA model the QE cross sections per
nucleon target of electron scattering on ${}^{40}$Ca (${}^{40}$Ar) are lower than
those for ${}^{12}$C. The effect of 15\% is observed in the QE region and  is
higher than experimental errors.

For electron scattering on the carbon and calcium targets we evaluated the
ratios of the calculated inclusive cross sections to the measured ones at the
momentum transfer $|\q|_{dip}$ that corresponds to the minimum of the measured
cross sections in the dip region. We also estimated the
$\mathrm{2p}$-$\mathrm{2h}$ MEC contribution
to the $(e,e')$ cross section at $|\q|_{dip}$. At the $|\q|_{dip}<250$ MeV the
RDWIA+MEC+RES approach underestimates the measured cross sections by about 30\%
and is in agreement with data within the experimental uncertainties at
$|\q|_{dip}\geq 500$ MeV. The MEC contribution decreases with
$|\q|_{dip}$ from 65\% at $|\q|_{dip}=250$ to 20\% at $|\q|_{dip}=1000$. These
results depend weakly on electron beam energy. So, we validated the
RDWIA+MEC+RES approach in the vector sector of the electroweak interaction by
describing ${}^{12}$C, ${}^{40}$Ca, and ${}^{40}$Ar data.

\section*{Acknowledgments}

The authors greatly acknowledge J. Amaro and G. Megias for fruitful 
discussions  and for putting in our disposal the codes for calculation of the 
MEC's electroweak response functions that were used in this work. We specially
thank R. Kokoulin and A. Habig for fruitful discussians and a critical reading
of the manuscript.

%



\begin{thebibliography}{99}

\bibitem{NOvA1} M.~A.~Acero {\it et al.}, (NOvA Collaboration), Phys. Rev. Lett.
{\bf 123}, 151803 (2019).
\bibitem{T2K} K.~Abe {\it et al.}, (T2K Collaboration), Phys. Rev. Lett.
{\bf 121}, 171802 (2018).
\bibitem{DUNE} R.~Acciarri {\it et al.}, (DUNE Collaboration), 
  FERMILAB-DESIGN-2016-03.
\bibitem{HK2T} K.~Abe {\it et al.}, (Hyper-Kamiokande Collaboration)
  arXiv:1805.04163 [physics.ins-det].
\bibitem{NOvA2} M.~A.~Acero {\it et al.}, (NOvA Collaboration), Phys. Rev. Lett.
{\bf D98}, 032012 (2018).
\bibitem{Katori} T.~Katori, M.~ Martini, J. Phys. {\bf G45}, 013001 (2018).
\bibitem{Alvares} L.~Alvarez-Ruso {\it et al.} Prog. Part. Nucl. Phys.
  {\bf 100}, 1 (2018). 
\bibitem{BAV1} A.~V.~Butkevich and S.~A.~Kulagin, Phys. Rev. {\bf C76},
  045502 (2007).
\bibitem{BAV2} A.~V.~Butkevich, Phys. Rev. {\bf C80}, 014610 (2009).
\bibitem{BAV3} A.~V.~Butkevich, Phys. Rev. {\bf C82}, 055501 (2010).
\bibitem{Blund} P.~G.~Blunden and M.~N.~Batler, Phys. Lett. {\bf B219}, 151
  (1989)  
\bibitem{Martini1} M.~Martini, M.~Ericson, and G.~Chanfray, 
Phys. Rev. {\bf C84}, 055502 (2011)
\bibitem{Martini2} M.~Martini, and M.~Ericson, 
Phys. Rev. {\bf C87}, 065501 (2013).
\bibitem{Nieves1} J.~Nieves, I.~Ruiz~Simo, and M.~J.~Vicente~Vacas, Phys. Lett. 
{\bf B707}, 72 (2012).
\bibitem{Nieves2} J.~Nieves, I.~Ruiz~Simo, and M.~J.~Vicente~Vacas, Phys. Lett. 
{\bf B721}, 90 (2013).
\bibitem{BAV4} A.~V.~Butkevich, Phys. Rev. {\bf C85}, 065501 (2012).
\bibitem{Martini3} M.~Martini, N.~Jachowicz, M.~Ericson, V.~Pandey,
  T.~Van~Cuyck, and N.~Van~Dessel, Phys. Rev. {\bf C94}, 015501 (2016).
\bibitem{Simo} I.~Ruiz~Simo, J.~E.~Amaro, M.~B.~Barbaro, A.~De~Pace, 
J.~A.~Caballero, and T.~W.~Donnelly, J. Phys. {\bf G44}, 065105 (2017).
\bibitem{Megias1} G.~D.~Megias, T.~W.~Donnelly, O.~Moreno,  
C.~F.~Williamson, J.~A.~Caballero, R.~Gonzalez-Jimenez, A.~De~Pace, 
M.~B.~Barbaro, W.~M.~Alberico, M.~Nardi, and J.~E.~Amaro, Phys. Rev. {\bf D91}, 
073004 (2015).
\bibitem{Megias2} G.~D.~Megias, J.~E.~Amaro, M.~B.~Barbaro, J.~A.~Caballero, 
T.~W.~Donnelly, Phys. Rev. {\bf D94}, 013012 (2016).
\bibitem{Megias3} G.~D.~Megias, J.~E.~Amaro, M.~B.~Barbaro, J.~A.~Caballero, 
T.~W.~Donnelly, and I.~R.~Simo, Phys. Rev. {\bf D94}, 093004 (2016).  
\bibitem{Rocco} Noemi~Rocco, Carlo~Barbieri, Omar~Benhar, Arturo~De~Pace, and
  Alessandro~Lovato, Phys. Rev. {\bf C99}, 025502 (2019).
\bibitem{BAV5} A.~V.~Butkevich and S.~V.~Luchuk, Phys. Rev. {\bf C97}, 045502
  (2018)
\bibitem{BAV6} A.~V.~Butkevich and S.~V.~Luchuk, Phys. Rev. {\bf D99}, 093001
  (2019)
\bibitem{Dolan} S.~Dolan, G.~D.~Megias, and S.~Bolognesi, Phys. Rev.
  {\bf D101}, 033003 (2020).
\bibitem{Gon1} M.~B.~Barbaro, J.~A.~Caballero, A.~De~Pace, T.~W.~Donnelly,
  R.~Gonzalez-Jimenez, G.~D.~Megias, Phys. Rev. {\bf C99}, 042501(R) (2019).
\bibitem{Gon2} R.~Gonzalez-Jimenez, A.~Nikolakopoulos, N.~Jachowicz,
  J.~M.~Udias, Phys. Rev. {\bf C100}, 045501 (2019).
\bibitem{Gon3} R.~Gonzalez-Jimenez, M.~B.~Barbaro, J.~A.~Caballero,
  T.~W.~Donnelly, N.~Jachowicz, G.~D.~Megias, K.~Niewczas, A.~Nikolakopoulos,
  J.~M.~Udias, Phys. Rev. {\bf C101}, 015503 (2020).
\bibitem{Pick}  A.~Picklesimer, J.~W.~Van Orden, S.~J.~Wallace,
Phys. Rev. {\bf C32}, 1312 (1985).
\bibitem{Udias} J.~M.~Udias, P.~Sarriguren, E.~Moya de Guerra, E.~Garrido, and
J.~A.~Caballero, Phys. Rev. {\bf C51}, 3246 (1995).
\bibitem{JKelly} James~J.~Kelly, Phys. Rev. {\bf C59}, 3256 (1999).
\bibitem{Bost1} P.~E.~Bosted and M.~E.~Christy, Phys. Rev. {\bf C77}, 065206
  (2008).
\bibitem{Bost2} M.~E.~Christy and P.~E.~Bosted, Phys. Rev. {\bf C81}, 055213
  (2010).
\bibitem{MMD} P.~Mergell, U.-G.~Meissner, and D.~Drechsel, Nucl. Phys.
{\bf A596}, 367 (1996).
\bibitem{deFor} T.~de~Forest, Nucl. Phys. {\bf A392}, 232 (1983).
\bibitem{Serot} B.~Serot, J.~Walecka, Adv. Nucl. Phys. {\bf 16}, 1 (1986).
\bibitem{Horow} C.~J.~Horowitz D.~P.~Murdock, and Brian~D.~Serot, in 
{\it Computational Nuclear Physics 1: Nuclear Structure} edited 
by K.~Langanke, J.~A.~Maruhn, Steven~E.~Koonin (Springer-Verlag,Berlin, 1991), 
p.129.
\bibitem{Dutta} D.~Dutta {\it et al.}, Phys. Rev. {\bf C68}, 064603 (2003).
\bibitem{Kelly1} J.~J.~Kelly, Phys. Rev. {\bf C71}, 064610 (2005).
\bibitem{Cooper} E.~D.~Cooper, S.~Hama, B.~C.~Clark, and R.~L.~Mercer,
Phys. Rev. {\bf C47}, 297 (1993).
\bibitem{Meucci1} A.~Meucci, C.~Giusti, and F.~D.~Pacati, Nucl. Phys.
{\bf A739}, 277 (2004).
\bibitem{Meucci2} A.~Meucci, C.~Giusti, and F.~D.~Pacati, Nucl. Phys.
{\bf A765}, 126 (2006).
\bibitem{Rocco2} N.~Rocco, L.~Alvarez-Ruso, A.~Lovato, and J. Nieves,
  Phys. Rev. {\bf C96}, 015504 (2017).
\bibitem{Atti} C.~Ciofi~degli~Atti and S.~Simula, Phys. Rev. {\bf C53 },
1689 (1996).
\bibitem{Pace} A.~De~Pace, M.~Nardi, W.~M.~Alberico, T.~W.~Donnelly, and 
A.~Molinari, Nucl. Phys. {\bf A726}, 303 (2003).
\bibitem{Her} E.~Hernandez, J.~Nieves, and M.~ Valverde, Phys. Rev. {\bf D76 },
033005 (2007).
\bibitem{Megias4} G.~D.~Megias, M.~B.~Barbaro, J.~A.~Caballero, J.~E.~Amaro,~
  T.~W.~Donnelly,~I.~Ruiz Simo,~J.~W~Van Orden, J. Phys. G{\bf 46}, 015104
  (2019)
\bibitem{MeAm} G.~D.~Megias and J.~E.~Amaro, Private communication.
\bibitem{Barreau} P.~Barreau  {\it et al.}, Nucl. Phys. {\bf A402} 515, (1983).
\bibitem{Whitney} R.~R.~Whitney, I.~Sick, J.~R.~Ficenec, R.~D.~Kephart, and
  W.~P.~Trower, Phys. Rev. {\bf C9}, 2230 (1974).
\bibitem{Connel} J.~S.~O'Connell, W.~R.~Dodge, J.~W.~Lightbody, Jr.,
  X.~K.~Maruyama, J.~O.~Adler, K.~Hansen, B.~Schroder, A.~M.~Bernstein,
  K.~I.~Blomqvist, B.~H.~Cottman, J.~J.~Comuzzi, R.~A.~Miskimen, B.~P.~Quinn,
  J.~H.~Koch, N.~Ohtsuka, Phys. Rev. {\bf C35} 1063 (1987).
\bibitem{Baran} D.~T.~Baran, B.~W.~Filippone, D.~Geesaman, M.~Green, R.~J.~Holt,
 H.~E.~Jackson, J.~Jourdan, R.~D.~McKeown, R.~G.~Milner, J.~Morgenstern, 
D.~H.~Potterveld, R.~E.~Segel, P.~Seidl, R.~C.~Walker, B.~Zeidman, 
Phys. Rev. Lett. {\bf 61}, 400 (1988).
\bibitem{Sealock} R.~M.~Sealock, K.~L.~Giovanetti, S.~T.~Thornton,
  Z.~E.~Meziani, O.~A.~Rondon-Aramayo, S.~Auffret, J.~P.~Chen, D.~G.Christian,
  D.~B.~Day, J.~S.~McCarthy, and R.~C.~Minehart, L.~C.~Dennis, K.~W.~Kemper,
  B.~A.~Mecking, J.~Morgenstern, Phys. Rev. Lett. {\bf 62}, 1350, (1989).
\bibitem{Benhar1} O.~Benhar, D.~Day, and I.~Sick, Rev. Mod. Phys. {\bf 80} 189 
 (2008).
\bibitem{Benhar2} O.~Benhar, D.~Day, I.~Sick, Rev. Mod. Phys. 
arXiv:nucl-ex/0603032 (2006).
\bibitem{William} C.~F.~Williamson, T.~C.~Yates, W.~M.~Schmitt, M.~Osborn,
  M.~Deady, Peter~D.~Zimmerman, C.~C.~Blatchley, Kamal~K.~Seth, M.~Sarmiento,
  B.~Parker, Yanhe~Jin, L.~E.~Wright, D.~S.~Onley, Phys. Rev. {\bf C56}, 3152
  (1997).
\bibitem{Dai1} H.~Dai  {\it et al.}, Phys. Res. {\bf C98} 014617 (2018).
\bibitem{Dai2} H.~Dai  {\it et al.}, Phys. Res. {\bf C99} 054608 (2019).
\bibitem{Anghi} M.~Anghinolfi  {\it et al.}, J. Phys. G.:{\bf 21}, L9 (1995).

\end{thebibliography}
\end{document}